\documentclass[fleqn,usenatbib]{mnras}

\usepackage[T1]{fontenc}

\DeclareRobustCommand{\VAN}[3]{#2}
\let\VANthebibliography\thebibliography
\def\thebibliography{\DeclareRobustCommand{\VAN}[3]{##3}\VANthebibliography}

\usepackage{graphicx}	
\usepackage{amsmath}	
\usepackage{amssymb}	

\def\hi{H\,{\sc i}}

\def\kms{km~s$^{-1}$}

\def\msun{M$_{\odot}$}

\def\fgas{$f_\mathrm{gas}$}
\def\mueff{$<\mu_\mathrm{eff}>$}
\def\jm{$j_*$~--~$M_*$}

\title[ALFALFA \jm\ relations]{On the existence of a tight planar relation between stellar specific angular momentum, mass and effective surface brightness for ALFALFA galaxies}

\author[E. Elson]{
E. Elson$^{1}$\thanks{E-mail: eelson@uwc.ac.za (EE)}
\\
$^{1}$Department of Physics \& Astronomy, University of the Western Cape, Robert Sobukwe Rd, Bellville, 7535, South Africa\\
}

\pubyear{2023}

\begin{document}
\label{firstpage}
\pagerange{\pageref{firstpage}--\pageref{lastpage}}
\maketitle

\begin{abstract}
Measurements of the dependence of stellar specific angular momentum ($j_*$) on stellar mass ($M_*$) are presented for large samples of ALFALFA galaxies spanning the stellar mass range $\sim 10^8$~--~$10^{11}$~\msun.  Accurate estimates of $j_*$ are generated using measurements of $I$-band effective radius and velocity width of the \hi\ line profile.  While the full sample ($N=3~607)$ of galaxies yields a \jm\ relation with power-law index $\alpha = 0.404\pm 0.03$, it is shown that various sub-samples have indices that are very similar to the best literature results, yet with comparatively lower intrinsic scatters.  A galaxy's mean $I$-band surface brightness within its effective radius (\mueff) is shown to significantly correlate with \jm\ scatter.  A 3D plane fit to all $N=3~607$ galaxies in $\log_{10}j_*$~-~$\log_{10}M_*$~-~$<\mu_\mathrm{eff}>$ space yields $j_*\propto M_*^{0.589\pm 0.002}<\mu_\mathrm{eff}>^{0.193\pm 0.002}$ with scatter $\sigma=0.089$~dex.   \mueff-selected sub-samples of size up to $N=1~450$ yield power-law \jm\ relations mostly consistent  with $\alpha=0.55\pm 0.02$ from the literature and with intrinsic scatter ranging from 0.083 to 0.129 dex.  Thus, this paper presents new, highly accurate measurements of the \jm\ relation that can be used to better understand the important roles played by angular momentum in the formation and evolution of galaxies. 
\end{abstract}



\section{Introduction}
Stellar mass and angular momentum play significant roles in the formation and evolution of galaxies.   40 years ago, the work of \citet{fall_1983} demonstrated the existence of a relationship between stellar mass ($M_*$) and stellar specific angular momentum ($j_*$).  In the \jm\ plane, galaxies are related via a power law, $j_*=\beta M_*^{\alpha}$ where the normalisation factor is related to galaxy morphology.  Today, this relation is called the Fall Relation.  From a theoretical perspective, the \jm\ relation is thought to be as a result of the ways in which structure growth in a $\Lambda$CDM universe is governed by the transfer of angular momentum from the tidal field of an irregular matter distribution to collapsing proto-galaxies (e.g., \citealt{peebles_1969}).  

In recent years, the \jm\ relation has been investigated in significant detail in the nearby Universe using a variety of galaxy samples and measurement techniques. \citet{romanowsky_fall_2012} and \citet{fall_romanowsky_2013} re-examined the role of angular momentum in a sample of $\sim 100$ nearby bright galaxies of all types spanning more than three orders of magnitude in stellar mass.  They showed spirals and ellipticals to follow parallel \jm\ tracks with log slopes of $\sim 0.6$.  They also demonstrated the way in which  a spiral galaxy's $j_*$ can be estimated to high accuracy using only global measurements (circular velocity and effective radius), which is the method adopted in this study. \citet{OG14} used high-quality rotation curves together with optical and \hi\ imaging to measure the angular momenta of stars and gas in 16 nearby THINGS spirals.  This study arguably pioneered the use of so-called ``high-precision'' angular momentum measurements\footnote{Based on radially integrating circular velocity and radial mass profiles.} and also showed bulge-to-total mass (a proxy for morphology) to significantly control the normalisation of the relation.  Thereafter, several authors attempted to extend the measurement of the \jm\ relation to lower masses (e.g., \citealt{chowdhury_2017, elson_2017}) with some claims of the relation being different for gas-rich dwarfs\footnote{These studies also considered the angular momentum contained in gas.}.  \citet{posti_2018b} used a sample of 92 galaxies with high-quality rotation curves as well as spatially-resolved \hi\ and infrared image sets to generate a \jm\ relation with slope $\alpha=0.55\pm 0.02$ and an orthogonal intrinsic scatter of $0.17\pm 0.01$~dex.  This \jm\ relation is generally regarded as being the most accurate one produced to date.  

While the \jm\ relation is known to depend fundamentally on morphological type, several authors have attempted to identify physical drivers of its scatter. For a sample of 488 galaxies extracted from the SAMI Galaxy survey, \citet{cortese_2016} showed the scatter in the \jm\ relation to be strongly correlated with optical morphology (as classified visually and according to S\'ersic index).  \citet{mancera_pina_2021b} reported the existence of tight correlations between $j$, $M$ and cold gas fraction of the interstellar medium for the stellar, gas and baryonic mass components of samples of up to 130 galaxies.  They showed the plane to be followed not only by typical disc galaxies, but also by galaxies with extreme mass and size properties which are outliers in standard 2D \jm\ relations.  Most recently, \citet{hardwick_2022b} found similar results for the baryonic specific angular momentum content of a sample of 559 galaxies with $M_*>10^9$~\msun\ from the xGASS survey, albeit with a larger scatter that they attribute to their sample covering a wide range of galaxy properties.  For a sample of 560 xGASS galaxies, \citet{hardwick_2022a} found \hi\ gas fraction is the strongest correlated parameter (to \jm\ scatter) for low stellar masses, while bulge-to-total mass ratio becomes slightly more dominant at higher masses.

Given that the shape of the \jm\ relation can be used to test galaxy formation models (e.g., \citealt{posti_2018b}), it is crucial to check whether it is well-described by an unbroken power law for large galaxy samples that span a range of morphologies.  To this end, the current study uses publicly-available data to study the \jm\ relation for ALFALFA galaxies.  Relations are presented that have power-law indices and normalisation factors that are consistent with the most accurate \jm\ relations from the literature, yet which typically have much lower intrinsic scatters and which are based on samples of significantly larger sizes.  

The layout of this paper is as follows.  Section~\ref{sec:data} describes the various catalogues from which the required observed and derived galaxy properties are sourced.  Sample selection criteria are presented in Section~\ref{sec:sample} while the method of measuring $j_*$ is discussed in Section~\ref{methods}.  Results for the full sample are presented in Section~\ref{sec:results} as well as the results for various sub-samples based on gas fraction and effective surface brightness.  A brief demonstration of the manner in which a well-measured \jm\ relation from this study can be used to constrain the $M_*$ dependence of the stellar-to-halo specific angular momentum relation is given in Section~\ref{sec:fj}.  Finally, the results are summarised in Section~\ref{sec:summary}.

\section{Data}\label{sec:data}
The observed and derived galaxy properties used in this study come from three cross-matched catalogues.  

The primary data source is the  ALFALFA Extragalactic \hi\ Source Catalogue \citep{haynes_2018} from which measurements of the velocity width of the \hi\ line profile, $W_\mathrm{50}$ in units of \kms, and adopted distance, $D$ in units of Mpc, are acquired.  \citet{haynes_2018} state $W_\mathrm{50}$ to be corrected for instrumental broadening.   For galaxies with $cz \ge 6000$~\kms\ \citet{haynes_2018} estimate distances as $cz/H_\mathrm{0}$, where $cz$ is the recessional velocity and $H_\mathrm{0}$ is the Hubble Constant, adopted to be 70~\kms~Mpc$^{-1}$.  For galaxies with $cz<6000$~\kms\ the local peculiar velocity model of \citet{masters_2005} is used.  Where available, published primary distances are adopted, as well as secondary distances from \citet{tully_2013}.  Distance errors are generated in a Monte Carlo fashion as described in \citet{jones_2018}.  

\citet{durbala_2020} provide SDSS identifications for nearly all of the galaxies from the 100\% ALFALFA survey, including $\sim$~12~000 that do not have SDSS spectroscopy.  They also derive absolute magnitudes and stellar masses for the galaxies using optical colours.  \citet{durbala_2020}  include in their catalogue additional stellar mass estimates from the GALEX-SDSS-WISE Legacy Catalog 2 (GSWLC-2, \citealt{salim_2016, salim_2018}) based on ultraviolet/optical/infrared SED fitting, as well as infrared photometry from unWISE \citep{mcGaugh_2015} and/or ultraviolet imaging from GALEX.  All of these stellar mass estimates are utilised in the current study.  For any galaxy that has more than one stellar mass estimate, the logarithm of the mean of the available stellar masses is used. 

The SDSS identifications from \citet{durbala_2020} are used in this work to query Data Release 15 of the SDSS.  For each galaxy, based on an exponential profile fit to its azimuthally-averaged $I$-band surface brightness profile, measurements of effective radius and apparent magnitude are obtained.  

\section{Galaxy sample}\label{sec:sample}
In order to create a suitable sample for which accurate estimates of $j_*$ can be produced, several cuts are applied to the 100\% ALFALFA survey sample.  Only \hi\ spectra designated as ``code 1'' by \citet{haynes_2018} are utilised.  These sources have well-matched signal characteristics between the two independent polarisations  observed by ALFALFA and a spatial extent at least as large as the beam.  It is further required in this work that these sources have relative distance and $W_{50}$ uncertainties smaller than 0.1 and a signal-to-noise ratio $S/N>6$.

Given that an \hi-selected sample such as ALFALFA is expected to consist mainly of disc-dominated systems, only those galaxies with an $I$-band light profile that is better fit by an exponential profile than a de Vaucouleurs profile are used.  Therefore, this study focuses on the disc components of disc-dominated systems.  Additionally, such galaxies must have a measured effective radius $\ge 1.5$~arcsec in order to ensure they are spatially well-resolved, and the relative error on this measurement must be smaller than 0.1.  Galaxies are required to have an axial ratio measurement in the \citet{durbala_2020} catalogue  and an associated inclination (see Eqn~\ref{eqn:incl}) greater than 30 degrees.  The final cut that is applied is on the relative uncertainty of a galaxy's $j_*$ estimate (discussed in the following section).  Because $j_*$ estimates become very unreliable at low stellar masses, only galaxies with a relative $j_*$ uncertainty less than 0.2 are used in this study.  The effects of this cut on the sample are discussed below. 

\begin{figure*}
	\includegraphics[width=2\columnwidth]{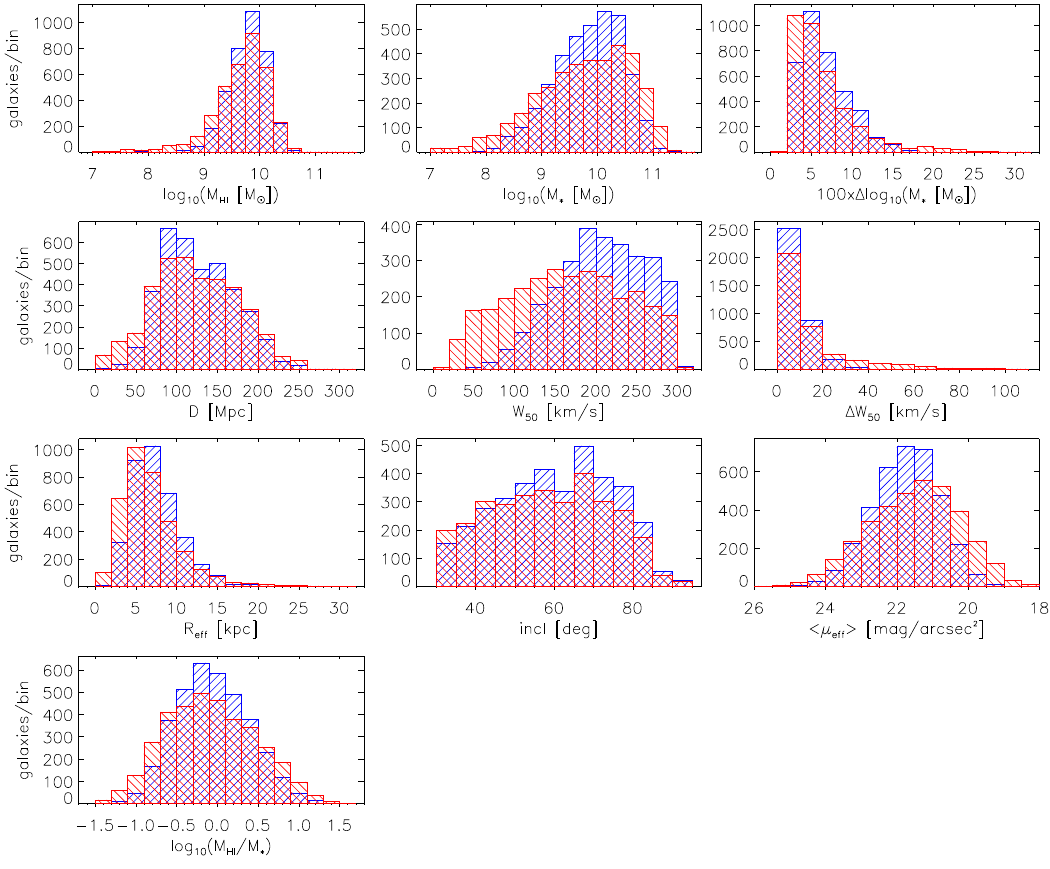}
    \caption{Distributions of various observed and derived properties of the ALFALFA galaxies.  Blue-hatched bins represent galaxies from the sample ($N=3~607$) used in this work, while red-hatched bins represented random subsets drawn from the full ALFALFA sample.  From left to right, top to bottom, the panels show the distributions of logarithm of the \hi\ mass, logarithm of the stellar mass, error in the stellar mass, distance (calculated as $cz/70$),  velocity width of the \hi\ line profile,  error in the velocity width, $I$-band effective radius from SDSS DR 15, disc inclination as calculated using Eqn~\ref{eqn:incl}, $I$-band effective surface brightness, and  logarithm of the \hi\ gas fraction.}
    \label{fig:gal_props}
\end{figure*}

Implementing all of the above-listed cuts results in a high-quality subset of 3~607 galaxies that serves as the primary sample for this study.   Figure~\ref{fig:gal_props} shows as blue-hatched histograms the distributions of the various above-listed observed and derived properties of these galaxies, as well as some additional properties.   Also shown (as red-hatched histograms) are the corresponding distributions for the parent ALFALFA sample \footnote{Random subsets of size $N=3~607$ are used for the sake of easy comparison.}.  

From Figure~\ref{fig:gal_props} it is clear that the cuts applied to the full ALFALFA data sample do not significantly alter the distributions of the various observed and derived quantities of the galaxy sample used in this study.  Perhaps the one exception is the $W_{50}$ distribution.  The reduced number of galaxies with  $W_{50} \lesssim 150$~\kms\ is as a result of the above-mentioned S/N and inclination cuts that preferentially remove low-mass dwarf systems.  In Section~\ref{methods}, the method used in this paper to generate  uncertainty estimates for $j_*$ is presented.  All galaxies with a relative error in their $j_*$ estimate greater that 0.2 are also removed to form the final sample of $N=3~607$ galaxies.  Such a cut also preferentially removes low-mass galaxies for which $j_*$ is difficult to estimate.





\section{Methods}\label{methods}
A general expression for the stellar specific angular momentum content of a flat galaxy with circular orbits is
\begin{equation}
j_*= \left|  {\int \vec{r} \times  \vec{v} dM \over \int dM} \right|.
\end{equation}

Given a measure of a galaxy's stellar rotation curve and azimuthally-averaged stellar mass surface density profile, $j_*$ can be measured very accurately.   However, in lieu of spatially-resolved image sets for a galaxy, an approximation of $j_*$ can be obtained using global measurements. If  a spiral galaxy's density distribution is approximated as an infinitely-thin exponential disc and if a radially-constant rotation curve is assumed, then an accurate approximation of its specific angular momentum is given by the expression
\begin{equation}\label{eqn:j}
\tilde{j_*} = 2 V_\mathrm{circ}R_\mathrm{d}, 
\end{equation}
where 
$V_\mathrm{circ}$ is the circular velocity and $R_\mathrm{d}$ is the exponential disc scale length.  \citet{romanowsky_fall_2012} explicitly show Eqn~\ref{eqn:j} to provide an excellent approximation for the true stellar specific angular momenta for real galaxies whose rotation curves vary with radius.  In this work, Eqn~\ref{eqn:j} is used to estimate $j_*$ for each of the $N=3~607$ galaxies making up the primary sample.  
 
The velocity width of the \hi\ line profile of each galaxy is combined with the measure of its disc inclination angle to calculate its maximum circular velocity: $V_\mathrm{circ} = W_\mathrm{50}/2\sin(i)$.  Estimates of disc inclination come from the axial ratios ($b/a$) in the \citet{durbala_2020} catalogue using
\begin{equation}\label{eqn:incl}
\cos^2i={(b/a)^2-q_0^2\over 1-q_0^2}, 
\end{equation}
where an intrinsic axial ratio of the disc of $q_0=0.15$ is adopted.  A galaxy's $R_\mathrm{d}$ measurement comes from the cross-matched SDSS DR15 data.  The catalogue actually offers a measure of the effective radius in units of arcsec, but this is easily converted to a scale length in units of kpc under the assumption of an exponential disc\footnote{$R_\mathrm{eff}=1.68R_\mathrm{d}$.} and by incorporating the galaxy's distance measurement from the ALFALFA catalogue.  

In order for the calculated stellar specific angular momenta to be used appropriately, associated uncertainty measurements are required.  In this work a very conservative approach is adopted in order to calculate these.  Given that a galaxy's $j_*$  is calculated from the measurements of its $W_\mathrm{50}$, disk inclination, $R_\mathrm{d}$ and distance, all combinations of the maximum and minimum values of these quantities (according to their quoted uncertainties from their source catalogues) are used to generate an additional 16 measurements of $j_*$.  The maximum difference between any two of these additional measurements is taken as the uncertainty, $\Delta j_*$, of $j_*$.  These propagated statistical uncertainties do not account for other sources of uncertainty that are likely present.  Systematic uncertainties will arise from various aspects of the dominant formation and evolutionary process.  Mass-dependent relationships between \hi\ line widths and circular velocities, mass-dependent deviations from exponential light profiles, rising or declining outer rotation curves, variable mass-to-light ratios, non-circular orbits of the \hi, etc. will all contribute to the total error.  Thus, the $j_*$ and $M_*$ uncertainties used in this study are considered lower estimates of the total uncertainties.

\begin{figure}
	\includegraphics[width=\columnwidth]{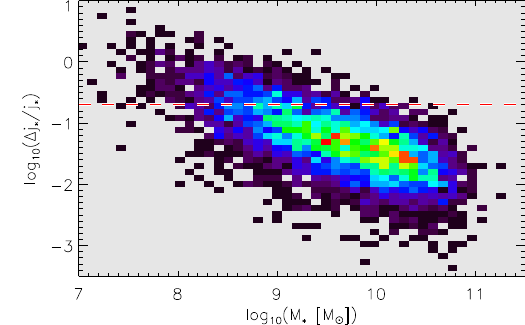}
    \caption{2D distribution of relative uncertainty in $j_*$ as a function of $M_*$. The horizontal dashed line marks a relative uncertainty $\Delta j_*/j_* =0.2$.  Galaxies above this uncertainty threshold are excluded from the present study.  Thus, the $N=3~607$ galaxies making up the primary sample are those distributed below the dashed line.  Their average relative uncertainty is $\Delta j_*/j_*\sim 0.05$.}
    \label{fig:jerrors}
\end{figure}

For the $N=3~607$ galaxies used in this study, Figure~\ref{fig:jerrors} shows the distribution of the relative uncertainties, $\Delta j_*/j_*$, as a function of stellar mass.  While the large majority of galaxies with $\log_{10}M_*/M_{\odot} \gtrsim 9$ have relative errors less than 0.1, galaxies below this mass threshold have significantly larger uncertainties.  The main source of this uncertainty is easily traced back to the uncertainty in the $I$-band effective radius acquired from SDSS DR15.  This is not surprising given that such a measurement is much more difficult to make for low-mass dwarf systems than it is for disc-dominated spirals.  The horizontal dashed line in Figure~\ref{fig:jerrors} indicates a relative uncertainty in $j_*$ of 0.2.  Galaxies with a higher uncertainty are not used in the present study.  Thus, the $N=3~607$ galaxies used in this study are those distributed below the horizontal dashed line.  The mean relative uncertainty on $j_*$ for these galaxies is $\Delta j_*/j_*\sim 0.05$.

\section{Results}\label{sec:results}
\begin{figure}
	\includegraphics[width=\columnwidth]{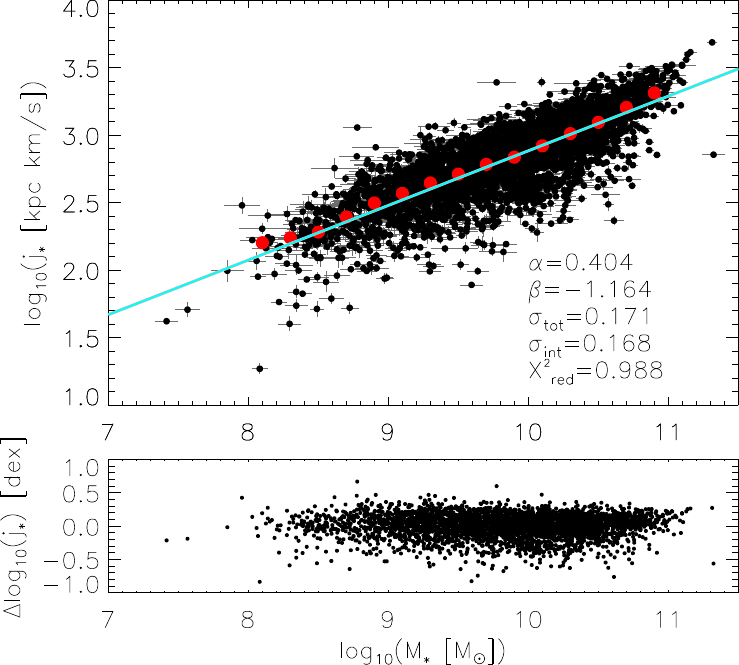}
    \caption{Top panel: Stellar specific angular momentum as a function of stellar mass for the full $N=3~607$ sample of ALFALFA galaxies considered in this study.  Grey error bars represent the uncertainties on $j_*$ and $M_*$ for each galaxy.  Red-filled circles represent the median $\log_{10}j_*$ in $\log_{10}M_*$ intervals of 0.2 dex.  The index ($\alpha$) and normalisation factor ($\beta$) of the best-fitting power-law model, represented by the overlaid cyan line, are given in the bottom right of the panel.  This model is fit directly to the black points (not the red-filled) circles by taking into account their uncertainties along each axis.  The intrinsic scatter associated with this model is $\sigma_\mathrm{int}=0.168$~dex while total scatter is $\sigma_\mathrm{tot}=0.171$~dex.  The reduced $\chi^2$ statistic for the best-fit model is $\chi^2_\mathrm{red}=0.988$.  Bottom panel: $j_*$ residuals as a function of stellar mass.  Clearly, there is no systematic trend with mass.}
    \label{fig:j-M}
\end{figure}

Figure~\ref{fig:j-M} shows $\log_{10}j_*$ as a function of $\log_{10}M_*$ for the full $N=3~607$ sample of ALFALFA galaxies utilised in this study.  The overlaid solid cyan line represents the best-fitting power-law model relating $j_*$ to $M_*$, given by 
\begin{equation}\label{eqn:j-M}
\log_{10}\left( {j_*\over\mathrm{kpc~km~s^{-1}}} \right) = \alpha\log_{10}\left({M_*\over M_\odot}\right) +\beta. 
\end{equation}
This model was fit using the MPFITEXY routine \citep{mpfitexy} which itself depends on the MPFIT package \citep{mpfit}.  MPFITEXY is able to incorporate the errors of both the x and y variables into the fit.  It also uses the procedure described in \citet{bedregal_2006} to  automatically adjust an estimate of the intrinsic scatter of the data in order to ensure the reduced $\chi^2$ statistic is very close to unity.  

For the full sample of $N=3~607$ ALFALFA galaxies, the best-fit power law index for the \jm\ relation is $\alpha=0.404\pm 0.003$ while the best-fit normalisation factor is $\beta=-1.164\pm 0.03$.  The uncertainty estimates  are generated using a Monte-Carlo procedure in which the power-law model is fit to 1000 randomly-selected 75\% subsets of the data.  The resulting distributions of the two parameters are well-modelled by Gaussians, and its their standard deviations that are used as the parameter uncertainty estimates\footnote{In this study, all parameter uncertainties are generated in this manner.}.  The intrinsic and total scatters of the data about the best-fit model are estimated to be 0.171 and 0.168~dex, respectively, while the reduced $\chi^2$ statistic is $\chi^2_\mathrm{red}=0.988$.  The estimated intrinsic scatter being so close to the total scatter likely indicates that the $j_*$ uncertainties calculated in this study represent propagated statistical uncertainties, and that they do not fully capture sources of systematic uncertainty.

Shown in the bottom panel of Figure~\ref{fig:j-M} is the difference between the $j_*$ estimates and the best-fit model (i.e., the $j_*$ residuals).  Very clear is the fact that there is no systematic trend of the residuals for the \jm\ relation as a function of stellar mass.  Within the scatter, a galaxy seems equally likely to be below or above the best-fit relation.  Results from all of the models fitted in this study are summarised in Table~\ref{table1}.

\begin{table*}
	\centering
	\caption{Best-fit parameters and statistics for all fitted models.  The top part of the table presents results for Eqn~\ref{eqn:j-M} fit to the full $N=3~607$ sample of ALFALFA galaxies, and then to various subsets selected by \fgas\ and then \mueff, all in 2D space.  The bottom part of the table show results for Eqns~\ref{eqn:j_M_fgas} and \ref{eqn:j_M_mueff} fit in 3D space.  From left to right, the columns are: sample size, sample selection criterion, best-fitting $\alpha$, $\beta$ and $\gamma$ parameters each with $1\sigma$ uncertainty estimate, intrinsic/total scatter of the data about the best-fit 2D/3D model, reduced $\chi^2$ statistic for the best-fit model.}
	\label{table1}
	\begin{tabular}{ccccccc} 
		\hline
		N & sample & $\alpha$ & $\beta$ & $\gamma$ & $\sigma$ & $\chi^2_\mathrm{red}$\\
		\hline
		\\
		\textbf{2D relations} & &&&&&\\
		\\
		3607 & all & $0.40\pm 0.003$ & $-1.16\pm 0.03$ &  --& 0.168 & 0.988\\
		\\
		\textbf{$f_\mathrm{gas}$ subsets} & &&&&&\\
		\\
		29 & $f_\mathrm{gas}\in [1, 1.5)$ & $0.63\pm 0.056$ & $-3.05\pm 0.485$ & -- & 0.211 & 0.995\\
		385 & $f_\mathrm{gas}\in [0.5, 1)$ & $0.51\pm 0.014$ & $-2.08\pm 0.127$ &  -- & 0.159 & 1.011\\
		1146 & $f_\mathrm{gas}\in [0, 0.5)$ & $0.50\pm 0.007$ & $-2.13\pm 0.075$ &  --& 0.152 & 1.017\\
		1450 & $f_\mathrm{gas}\in [-0.5, 0)$ & $0.53\pm 0.007$ & $-2.47\pm 0.072$ &  --& 0.150 & 0.992\\
		597 & $f_\mathrm{gas}\in [-1, -0.5)$ & $0.70\pm 0.014$ & $-4.37\pm 0.153$ &  --& 0.157 & 1.009\\
		\\
		\textbf{$<\mu_\mathrm{eff}>$ subsets} & &&&&&\\
		\\
		36 & $<\mu_\mathrm{eff}> \in [24,25)$ & $0.61\pm 0.019$ & $-3.63\pm 0.195$ &  --& 0.108 & 1.010\\
		312 & $<\mu_\mathrm{eff}> \in [23,24)$ & $0.64\pm 0.007$ & $-3.75\pm 0.076$ &  --& 0.129 & 0.998\\
		1036 & $<\mu_\mathrm{eff}> \in [22,23)$ & $0.57\pm 0.004$ & $-2.87\pm 0.040$ &  --& 0.100 & 1.004\\
		1450 & $<\mu_\mathrm{eff}> \in [21,22)$ & $0.54\pm 0.004$ & $-2.37\pm 0.042$ &  --& 0.097 & 0.998\\
		694 & $<\mu_\mathrm{eff}> \in [20,21)$ & $0.57\pm 0.008$ & $-2.55\pm 0.075$ &  --& 0.083 & 1.015\\
		79 & $<\mu_\mathrm{eff}> \in [19,20)$ & $0.52\pm 0.029$ & $-2.00\pm 0.258$ &  --& 0.104 & 1.002\\
		\\
		\hline
		\\
		\textbf{3D relations} & &&&&&\\ 
		\\
		\textbf{$j_*$, $M_*$, \fgas} & &&&&&\\
		3607 & all & $0.56\pm 0.004$ & $0.276\pm 0.006$ & $-2.741\pm 0.045$ & 0.134 & 2.547\\
		\\
		\textbf{$j_*$, $M_*$, \mueff} & &&&&&\\
		3607 & all & $0.58\pm 0.002$ & $0.193\pm 0.002$ & $-7.198\pm 0.054$ & 0.089 & 1.154\\
		\\
		\hline
	\end{tabular}
\end{table*}

This best-fit power-law for the full sample of ALFALFA galaxies is perhaps most directly comparable to those of the investigation recently carried out by \citet{hardwick_2022a},  who study the \jm\ relation using a sample of 564 nearby galaxies from the eXtended GALEX Arecibo SDSS Survey.  As done in the current study, they use measurements of the velocity width of the \hi\ line to estimate each galaxy's maximum circular velocity.  Stellar mass and effective radius measurements come from S\'ersic profiles fitted to SDSS $I$-band images.  For their full sample that treats both the disc and bulge components to be co-rotating with the \hi, \citet{hardwick_2022a} measure a slope of $\alpha=0.47\pm0.02$ with a scatter of 0.22~dex of the data about the best-fit relation.  This is similar, yet not consistent with the results presented above.  However, the ALFALFA sample from the present study is much larger than the \citet{hardwick_2022a} sample and spans an additional $\sim 1.5$~decades of stellar masses below $10^{9.5}$~M$_{\odot}$.  Furthermore, the scatter about the best-fit relation is significantly smaller.  

Rather than delve further into the details of the \jm\ relation presented in Figure~\ref{fig:j-M}, an important point noted by \citet{hardwick_2022a} needs to be considered.  They note the fact that the \jm\ relation is known to vary with morphological type (e.g., \citealt{fall_1983, romanowsky_fall_2012}) and with properties that serve as proxies for galaxy morphology, such as bulge-to-disk ratio (e.g., \citealt{romanowsky_fall_2012, OG14}), gas fraction (e.g., \citealt{mancera_pina_2021b, hardwick_2022a}), stellar-light distribution (e.g., \citealt{cortese_2016}).  From Figure~\ref{fig:gal_props} that shows the distributions of stellar mass, \hi\ gas fraction and \hi\ line velocity width, it is clear that the full sample of $N=3~607$ ALFALFA galaxies spans a large range of galaxy type - from low-mass, gas-rich dwarfs to high-mass systems containing relatively little \hi.  Hence, in the sections that follow, the \jm\ relations for various sub-samples selected by gas fraction and then mean effective surface brightness are presented, parameterised, discussed, and compared to literature results. 

\subsection{The $j_*$~--~$M_*$~--~$f_\mathrm{gas}$ plane}
Several authors have explored whether residuals in the \jm\ relation correlate with a third parameter.  A particularly important driver of scatter in the stellar relation seems to be the cold gas fraction of the interstellar medium, $f_\mathrm{gas}$.  For late-type galaxies, this quantity is known to be well-correlated with stellar mass in the sense that low-mass systems have high gas fractions (oftentimes having their baryonic mass dominated by $M_\mathrm{HI}$) while high-mass systems typically have low gas fractions (e.g., \citealt{huang_2012}).  It is therefore reasonable to expect a correlation between \jm\ scatter and \fgas.

Using high-quality rotation curves together with HI and near-infrared surface density profiles for a sample of 130 galaxies, \citet{mancera_pina_2021a} reported that at fixed stellar mass, galaxies with higher \fgas\ typically also have higher $j_*$.  For the same sample of galaxies \citet{mancera_pina_2021b} reported the discovery of a 3D plane relating the logarithms of $j_*$, $M_*$ and \fgas, to which they fitted the model 
\begin{equation}\label{eqn:j_M_fgas}
\log_{10}\left( {j_*\over\mathrm{kpc~km~s^{-1}}} \right) = \alpha\log_{10}\left({M_*\over M_\odot}\right) +\beta\log_{10}(f_\mathrm{gas})+\gamma,
\end{equation}
and noted the orthogonal scatter of their best-fit plane ($\sigma=0.10\pm 0.01$~dex) to be significantly smaller than for their 2D relations. Given the similar ways in which $j_*$ residual is related to bulge-to-total mass fraction (e.g., \citealt{fall_1983, romanowsky_fall_2012, cortese_2016}), \citet{mancera_pina_2021b} suggest such relations to be the result of an underlying, common physical process.   

To investigate the driver/s of \jm\ scatter for the ALFALFA galaxies considered in the present study, Eqn~\ref{eqn:j_M_fgas} is fit to the full sample of $N=3~607$ galaxies.  Gas fraction is defined as the ratio of a galaxy's \hi\ mass to its stellar mass: $f_\mathrm{gas}\equiv M_\mathrm{HI}/M_*$.  Note that \citet{mancera_pina_2021a} included a Helium contribution such that $f_\mathrm{gas}= 1.33M_\mathrm{HI}/M_*$.  The best-fit model parameters are $\alpha=0.566\pm 0.004$, $\beta=0.276\pm0.006$, $\gamma=-2.741\pm 0.045$.  The dependence on mass is significantly higher than for the 2D case, likely due to the manner in which the relation is now essentially being modelled as function of galaxy type (i.e., \fgas).  The dependence of $\log_{10}j_*$ on $\log_{10}f_\mathrm{gas}$ is approximately half of what it is on $\log_{10}M_*$.  The total scatter of the data about the fitted plane is well-approximated by a Gaussian with standard deviation $\sigma= 0.134$~dex, hence slightly lower than what was found for the 2D relation.  For comparison, \citet{mancera_pina_2021b} obtained $\alpha=0.67\pm 0.03$, $\beta=0.51\pm0.08$, $\gamma=-3.62\pm 0.23$ with a $1\sigma$ intrinsic scatter of $0.10\pm 0.01$~dex.   The relative dependence of their model on $\beta$ is significantly higher than the model fitted here. 

\begin{figure*}
	\includegraphics[width=2\columnwidth]{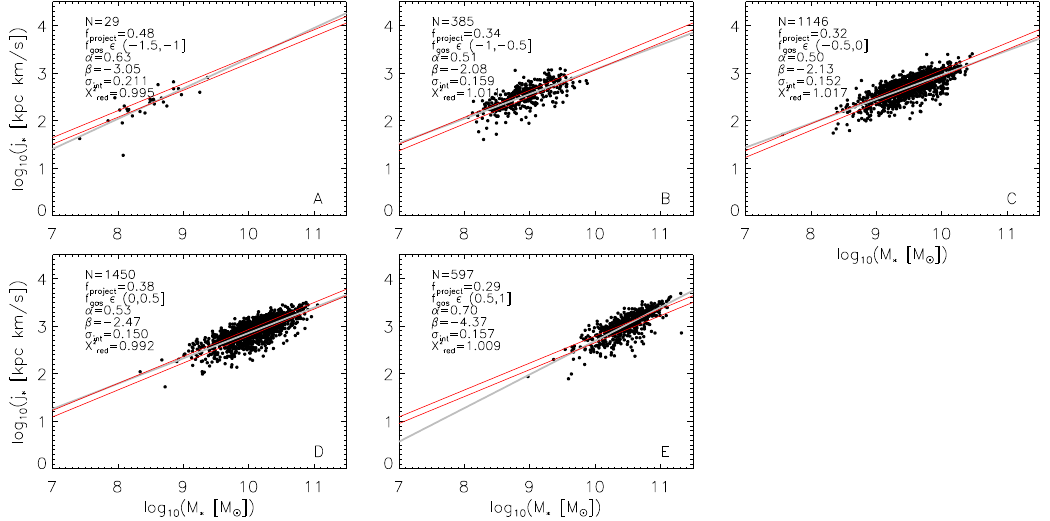}
    \caption{\jm\ relations for \fgas-selected sub-samples of the full $N=3~607$ sample of galaxies used in this study.  In each panel, the overlaid grey line represents the best-fitting power-law model (i.e., Eqn~\ref{eqn:j}) while the red lines represents the projections onto  2D \jm\ space of the 3D plane fitted to the galaxies in $\log_{10}j_*$~-~$\log_{10}M_*$~-~$\log_{10}f_\mathrm{gas}$ space.  Each projection shows only the portion of the 3D plane corresponding to the \fgas\ range corresponding to the panel.  While some galaxies are well-contained within the projection of the 3D plane, the majority are not.   In each panel, the following information is provided: number of galaxies in sub-sample, fraction of those galaxies that lie within the projection of the 3D plane, \fgas\ range of the galaxies, best-fitting power-law index, best-fitting power-law normalisation factor, intrinsic scatter of the data, reduced $\chi^2$ statistic of the best-fit model.}
    \label{fig:j-M-fgas}
\end{figure*}

Figure~\ref{fig:j-M-fgas} shows the 2D \jm\ relations for  \fgas-selected sub-samples of the $N=3~607$ ALFALFA galaxies.  The sub-samples are delimited by decreasing gas fractions of $\log_{10}f_\mathrm{gas}=\{1.5, 1.0, 0.5, 0.0, -0.5\}$.  All best-fit parameters and statistics are shown in Table~\ref{table1}.  Shown as a solid grey line in each panel of Figure~\ref{fig:j-M-fgas} is the best-fit model (i.e., Eqn~\ref{eqn:j}, again fit using MPFITEXY).  The solid red lines represent the projection of the fitted 3D model (i.e., Eqn~\ref{eqn:j_M_fgas}) onto the 2D \jm\ plane\footnote{For the corresponding $\log_{10}f_\mathrm{fgas}$ range.}.  Also given in each panel is the sample size, the fraction of galaxies lying within the projection of the fitted 3D plane, the standard deviation of the intrinsic scatter and the reduced $\chi^2$ statistic.  

The most noticeable difference between the \jm\ relations of the \fgas-selected sub-samples and that of the full galaxy sample is their higher power-law indices. The three sub-samples spanning $-0.5\le f_\mathrm{gas} <1.0$ (i.e., panels C, D, E)  have very similar power-law indices that are much closer to the value of $\alpha=0.56\pm 0.004$ for the fitted 3D plane.  Furthermore, these power-law indices are very close to the $\alpha=0.55\pm 0.02$ index of the best-fitting power law measured by \citet{posti_2018b} for a sample of 92 nearby galaxies with high-quality rotation curves and infrared photometry from the \textit{Spitzer} Photometry and Accurate Rotation Curves (SPARC, \citealt{SPARC}) sample.  Given the large stellar mass range spanned by their data set, as well as the very small intrinsic scatter of their \jm\ relation, theirs is generally considered to be the most accurate measurement of the relation to date. 
 
The results presented in Figure~\ref{fig:j-M-fgas} are significant in the sense that 1) the majority of the \fgas-selected sub-samples each contains many more galaxies that any other galaxy sample used to study the \jm\ relation to date, and 2) the best-fit power-law indices are very similar to those measured for samples of spatially-resolved galaxies with accurate rotation curves.  However, it is clear that considering \fgas\ as an additional parameter to which $j_*$ is correlated does not significantly reduce the scatters of the 2D relations. The lowest intrinsic scatter is $0.15$~dex, which is only slightly less than the intrinsic scatter of $0.168$~dex for the full $N=3~607$ sample.   This inability of \fgas\ to account for the residuals is further indicated by the fact that all of the 2D \jm\ relations based on \fgas-selected sub-samples have the \emph{minority} of their galaxies lying within their 2D projection of the best-fitting 3D plane.  Thus, in the following section, \fgas\ (a derived quantity) is replaced by a different (directly observable) galaxy property.

\begin{figure*}
	\includegraphics[width=2\columnwidth]{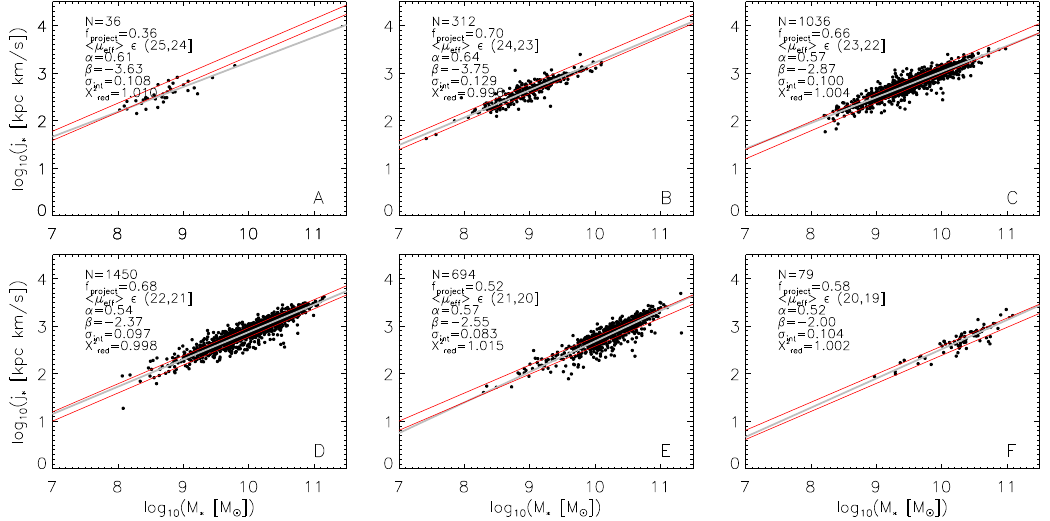}
    \caption{\jm\ relations for \mueff-selected sub-samples of the full $N=3~607$ sample of galaxies used in this study.  In each panel, the overlaid grey line represents the best-fitting power-law model (i.e., Eqn~\ref{eqn:j}) while the red lines represents the projections onto  2D \jm\ space of the 3D plane fitted to the galaxies in $\log_{10}j_*$~-~$\log_{10}M_*$~-~$<\mu_\mathrm{eff}>$ space.  Each projection shows only the portion of the 3D plane corresponding to the \mueff\ range corresponding to the panel.  Quite different from the results presented in Figure~\ref{fig:j-M-fgas} is the fact that most of the galaxies are now well-modelled by the 3D plane.  \mueff\ seems to be a much stonger influencer of a galaxy's $j_*$ content than \fgas\ is.  In each panel, the following information is provided: number of galaxies in sub-sample, fraction of those galaxies that lie within the projection of the 3D plane, \mueff\ range of the galaxies, best-fitting power-law index, best-fitting power-law normalisation factor, intrinsic scatter of the data, reduced $\chi^2$ statistic of the best-fit model.}
    \label{fig:j-M-mueff}
\end{figure*}

\subsection{The $j_*-M_*-<\mu_\mathrm{eff}>$ plane}
Several studies have demonstrated the dependence of the scatter in the \jm\ relation on a measure of central mass concentration such as bulge-to-total mass ratio, $\beta_*$.  The sample of 16 late-type galaxies presented by \citet{OG14} spanned a bulge fraction range of $0\le\beta_*\le 0.3$, while 90\% of the 92 galaxies from the \citet{posti_2018b} study spanned the same range. These samples therefore did not span the full $\beta_*$ range.  Part of the study presented by \citet{sweet_2018} is based on a sample of galaxies from the Calar Altar Legacy Integral Field Area (CALIFA) survey spanning a larger range of $0\le\beta_*\le 0.7$.  However, the most comprehensive studies of the importance of $\beta_*$ are arguably those of \citet{romanowsky_fall_2012} and \citet{fall_romanowsky_2013} whose samples consisted of 94 galaxies spanning the full range of bulge fraction, $0\le\beta_*\le 1$, with median $\beta_*=0.35$.  Taking bulge fraction into account, they found that both disk-dominated galaxies and bulge-dominated galaxies obey $j_*\propto M_*^{\alpha}$ relations with essentially the same exponent, $\alpha=0.6\pm0.1$, and with normalisation factors that differ by a factor~5. \citet{fall_romanowsky_2018} went on to show that $j_*$, $M_*$ and $\beta_*$ for normal galaxies of all morphological types are consistent with a model based on a linear superposition of independent disc and bulge components. 

Given the clearly important role played by a galaxy's stellar mass distribution in determining its location in \jm\ space, this works seeks to explore a similar possibility for the ALFALFA galaxies.  However, rather than consider $\beta_*$, a more accessible observable property is considered: the average $I$-band surface brightness of a galaxy within its $I$-band effective radius, \mueff.  The information extracted from the SDSS DR15 database is used to calculate this effective surface brightness quantity, the distribution of which for all $N=3~607$ galaxies is shown in Figure~\ref{fig:gal_props}.  A 3D plane is fit to the distribution of galaxies in $\log_{10}j_*$~-~$\log_{10}M_*$~-~$<\mu_\mathrm{eff}>$ space:
\begin{equation}\label{eqn:j_M_mueff}
\log_{10}\left( {j_*\over\mathrm{kpc~km~s^{-1}}} \right) = \alpha\log_{10}\left({M_*\over M_\odot}\right) +\beta\left({<\mu_\mathrm{eff}>\over \mathrm{mag~arcsec^{-2}}}\right)+\gamma.
\end{equation}
The best-fitting parameters are $\alpha=0.589\pm 0.002$, $\beta=0.193\pm 0.002$, $\gamma=-7.198\pm 0.054$.  Thus, this determination of $\alpha$ is consistent (within $3\sigma$) with that of the best-fit $\log_{10}j_* - \log_{10}M_* - \log_{10}f_\mathrm{gas}$ plane.  However, the most striking aspect of this model is the significantly lower $\log_{10}j_*$ residuals it offers.  These residuals are very well-approximated by a Gaussian with standard deviation of only $\sigma=0.089$~dex, making it much tighter than the best-fitting $\log_{10}j_* - \log_{10}M_* - \log_{10}f_\mathrm{gas}$ plane presented above, as well as the plane from of \citet{mancera_pina_2021b}.  It is worth noting a useful implication of the quality of fit of this plane to the data: it can be used to easily estimate a galaxy's stellar mass.  Given a measure of a system's \hi\ line width as well as optical or infrared imaging that allows for its effective radius and effective surface brightness to be measured, a fairly accurate estimate of its stellar mass can be inferred from Eqn~\ref{eqn:j_M_mueff}.

While the 3D plane tightly fits the data, fits of very similar quality are also found for the 2D \jm\ relations based on \mueff-selected sub-samples.  Figure~\ref{fig:j-M-mueff} shows the \jm\ relations for galaxies delimited by $<\mu_\mathrm{eff}>=\{25, 24, 23, 22, 21, 20, 19\}$~mag~arcsec$^{-2}$.  For each relation, the solid grey line is the best-fit  model (i.e., Eqn~\ref{eqn:j-M}).  The red lines delimited the projection of the best-fitting $\log_{10}j_* - \log_{10}M_*-<\mu_\mathrm{eff}>$ plane presented above onto the 2D \jm\ space\footnote{For the corresponding \mueff\ range.}. Also shown in each panel is the sample size, the fraction of galaxies that lie within the projection of the 3D plane (i.e., those between the red lines), the best-fit $\alpha$ and $\beta$ parameters, the standard deviation of the intrinsic $\log_{10}j_*$ scatter, and the reduced $\chi^2$ statistic.  

Immediately impactful are the relations shown in panels C, D and E corresponding to galaxies with effective surface brightness 23 - 20~mag~arcsec$^{-2}$. These all have $\alpha$ parameters consistent with the $\alpha=0.55\pm 0.02$ result from \citet{posti_2018b}.  Their normalisation factors are also all very close to that of the \citet{posti_2018b} model.  Those authors fit to their data a slightly different version of a power-law relation.  Suitably adjusting the normalisation constants for the relations from panels C, D, E yields values of $3.400\pm 0.040$, $3.570\pm 0.042$, $3.720\pm 0.075$, respectively, which can be compared to the value of $3.34\pm 0.03$ from the \citet{posti_2018b} study.  While the 92 galaxies making up the \citet{posti_2018b} sample do better probe the $8\le \log_{10}(M_*/M_{\odot})\le9$ mass range than the ALFALFA galaxies do, the ALFALFA relations exhibit  less intrinsic scatter for samples that are much larger. Best-fit models with intrinsic scatters of 0.100, 0.097 and 0.083~dex are achieved for sample sizes of 1036, 1450 and 694 galaxies, respectively. Furthermore, each of the 2D \jm\ relations shown in panels C, D, E have the large majority of their points lying within the corresponding projection of the best-fitting 3D $\log_{10}j_*$~-~$\log_{10}M_*$~-~$<\mu_\mathrm{eff}>$ plane.  The \jm\ relations shown in panels A and B are based on the sub-samples that have the lowest median stellar masses.  Their best-fitting power-law indices of $\alpha=0.61\pm 0.019$ and $\alpha=0.64\pm 0.007$ are closer to the value of 2/3 expected for dark matter haloes in a $\Lambda$CDM cosmology.  They are also consistent with several of the best-fit power-laws presented in Table~2 of \citet{romanowsky_fall_2012}  for various sub-samples of their $\sim100$ nearby bright galaxies of all types.  Yet, again, the scatters in the ALFALFA relations are significantly smaller. 

\begin{figure*}
	\includegraphics[width=2\columnwidth]{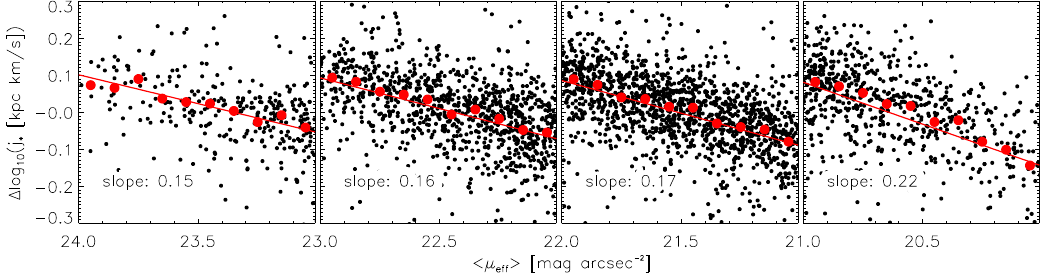}
    \caption{$\log_{10}j_*$ residual as a function of \mueff\ for the \jm relations shown in panels B, C, D, E of Figure~\ref{fig:j-M-mueff}.  Red-filled circles represent the mean residuals in \mueff\ bins of width 0.1~mag~arcsec$^{-2}$.  The solid red lines are linear fits to the individual galaxies (i.e., the black data points).  The slope of each fitted line is shown in the bottom left of each panel.  Note that the x-axes are arranged in a way that has the amount of \emph{flux} per unit area increasing from left to right.  Very clear is the dependence of $j_*$ residual on \mueff.  At a particular  $M_*$, a galaxy with higher/lower stellar flux per unit area within its effective radius will have less/more than the average amount of stellar specific angular momentum of galaxies at that mass.}
    \label{fig:delta_j_mueff}
\end{figure*}

To further demonstrate the dependence of $j_*$ on \mueff, Figure~\ref{fig:delta_j_mueff} shows $\log_{10}j_*$ residuals as a function of \mueff\ for the \jm\ relations shown in panels B, C, D, E of Figure~\ref{fig:j-M-mueff}.  Very clear from each of the panels in Figure~\ref{fig:delta_j_mueff} is the fact that the residuals are correlated with \mueff.  At a particular  $M_*$, a galaxy with higher/lower stellar flux per unit area within its effective radius will have less/more than the average amount of stellar specific angular momentum of galaxies at that mass.  This result is, of course, expected given the abundance of evidence in the literature for the role played by a galaxy's bulge-to-total mass ratio on its $j_*$ content, yet \mueff\ is a much simpler quantity that can be easily derived straight from photometric imaging.  In each panel of Figure~\ref{fig:j-M-mueff}, the large red-filled circles represent the mean $j_*$ residual over a  \mueff\ interval of 0.1~mag~arcsec$^{-2}$ while the line in each panel represents a first-order polynomial fit directly to the individual galaxies (not to the red-filled circles).   Clearly, the slope of the fitted line increases gradually as mean \mueff\ decreases (i.e., as the galaxies become brighter within their effective radii).  In other words, the effect of \mueff\ on a galaxy's $j_*$ content seems to be more pronounced for galaxies with denser concentrations of central starlight. 

If effective surface brightness, like other measures of mass concentration such as bulge-to-total mass fraction, is another observable quantity that significantly correlates with a galaxy's stellar specific angular momentum content, the reason must be linked to the manners in which galaxies have their mass distributions affected by various evolutionary processes.  While the ALFALFA population uniformly includes late-type galaxies that are more gas-rich for a given $M_*$ (e.g., \citealt{huang_2012}), high-mass galaxies will still typically have higher central mass concentrations than low-mass systems.  In terms of secular processes, high-mass spirals often have higher star formation rates which can contribute to a prominent central bulge.  Low-mass systems, however,  lack the gravitational potential required to sustain the intense star formation, and therefore have lower central mass concentrations.  The formation of a bar, which is statistically more prominent in massive spirals compared to low-mass spirals (e.g., \citealt{masters_2011, cervantes_sodi_2017}), can affect a galaxy's potential in ways that have gas clouds transported towards the central regions, thereby promoting star formation and stellar mass accumulation.  Gravitational resonances associated with bars can also lead to significant amounts of stellar mass being trapped in resonant orbits near the centres of galaxies.  Non-secular processes include mergers and interactions.  High-mass systems are likely to have experienced significant merger and/or interaction activity in recent times.  Such interactions are well-known to drive gas towards the central regions of a galaxy (e.g. \citealt{combes_1993}), thereby enhancing star formation and leading to the buildup of a larger bulge.  Any process that results in a galaxy's mass being more centrally-concentrated automatically reduces the size of the radius lever arm in $\vec{r}\times\vec{v}$.  Thus, the \jm\ relation is a sensitive probe of the evolutionary processes that ultimately determine the kinematics and mass distributions of galaxies. 

In summary, the \jm\ relations  based on \mueff-selected sub-samples of ALFALFA galaxies are among the tightest produced to date, and are based on sample sizes that are up to  a factor $\sim 3$ greater than any other sample presented in the literature, and up to a factor $\sim 15$ greater than the samples used for the most impactful \jm\ studies published so far.

\section{Constraining the $\MakeLowercase{f_j}$~-~$M_*$ connection}\label{sec:fj}
Accurate measurements of the \jm\ relation such as those presented in this work can be used to better constrain the links between the properties of dark matter haloes and the galaxies residing within them.  In $\Lambda$CDM theory, the stellar discs of galaxies are generally treated as having an amount of specific angular momentum that is some fraction of the specific angular momentum of their halo.  Early galaxy formation models (e.g., \citealt{fall_efstathiou_1980, mo_1998}) mostly assumed a direct proportionality relationship, and the same for the dependence of stellar mass on halo mass.  However, more recent results (e.g., \citealt{moster_2013, behroozi_2013}) have shown the stellar-to-halo mass relation (SHMR) to be very complex and non-linear.  \citet{posti_2018a} present a method of constraining the stellar-to-halo specific angular momentum relation (SHSAMR) for an assumed SHMR while also requiring the empirical \jm\ relation to be reproduced.   In this section, their procedure is emulated with the intention of demonstrating how the uncertainty in the SHSAMR can be reduced using the tight \jm\ relations presented in this work. 

In their Eqn~8, \citet{posti_2018a} present an expression for a dark matter halo's specific angular momentum:
\begin{equation}\label{eqn:jh}
j_\mathrm{h}={1.67\times 10^3\over \sqrt{F_E(c)}} \left({\lambda\over 0.035}\right)\left({M_\mathrm{h}\over 10^{12}M_\mathrm{\odot}}\right)^{2/3}~\mathrm{kpc~km~s^{-1}},
\end{equation}
where $\lambda$ is the spin parameter used to quantify the rotation acquired by a dark matter halo from tidal torques exerted by the surrounding matter (see \citealt{peebles_1969}), and $F_E(c)$ is a dimensionless factor that depends on the structure of the halo and which can be written as a function of its concentration parameter (see \citealt{posti_2018a} Eqn~7).   In their Eqn~9, they then give the expression for a galaxy's stellar specific angular momentum:
\begin{equation}\label{eqn:jstar}
j_*={77.4\over \sqrt{F_E(c)}} \left({\lambda\over 0.035}\right)f_jf_*^{-2/3}\left({M_*\over 10^{10}M_\mathrm{\odot}}\right)~\mathrm{kpc~km~s^{-1}},
\end{equation}
where $f_j\equiv j_*/j_\mathrm{h}$ is the ratio of the average specific angular momentum in stars to that of the halo, while $f_*\equiv M_*/M_\mathrm{h}$ is the ratio of stellar mass to halo mass.  These equations are used below together with the tightest \jm\ relation presented in this work (i.e., the one presented in panel E of Figure~\ref{fig:j-M-mueff}, with intrinsic scatter $\sigma_\mathrm{int}=0.083$~dex).

Guided by the steps outlined in \citet{posti_2018a}, the procedure is as follows:
\begin{enumerate}
\item Draw a sample of 10$^6$ uniformly distributed samples of $\log_{10}M_*$ in the range $8\le \log_{10}(M_*/M_{\odot})\le 12$. 
\item To each stellar mass, assign a halo mass based on a selected SHMR.  For this work, the SHMR relation from \citet{moster_2013} is used:
\begin{equation}\label{moster}
{M_*\over M_\mathrm{h}}= 2N\left[\left({M_\mathrm{h}\over M_1}\right)^{-\delta}  + \left({M_\mathrm{h}\over M_1}\right)^{\gamma} \right], 
\end{equation} 
where $N$ is a normalisation factor and $M_1$ is a characteristic mass below which ${M_*/M_\mathrm{h}}$ increases with $M_\mathrm{h}$ (halo mass) and above which it decreases.  These rates of increase and decrease are specified by $\delta$ and $\gamma$, respectively.  In their Eqns~11-14, \citet{moster_2013} specify the redshift dependence of $\log_{10}M_1$, $N$, $\delta$ and $\gamma$ in terms of a set of fitted parameters $M_{10}$, $M_{11}$, $N_{10}$, $N_{11}$, $\beta_{10}$, $\beta_{11}$, $\gamma_{10}$, $\gamma_{11}$.  However, given the low redshift spanned by ALFALFA galaxies ($z<0.06$), $z=0$ is assumed for all of the fitted parameters, resulting in $\log_{10}M_1=M_{10}$, $N=N_{10}$, $\delta=\beta_{10}$ and $\gamma=\gamma_{10}$.   Eqn~\ref{moster} thus allows each $\log_{10}M_*$ sample to be paired with a sample of $\log_{10}M_\mathrm{h}$ to which a uniform scatter of 0.2 dex is added. 
\item Draw a random spin parameter from a normal distribution in $\log_{10}\lambda$ with mean 0.035 and standard deviation 0.025.  As \citet{posti_2018a} point out, cosmological simulations have found $\lambda$ to be log-normally distributed and independent of halo mass. The mean and standard deviation statistics adopted here come from \citet{maccio_2007}.
\item Draw a random value for the concentration parameter from a normal distribution in $\log_{10}c$ with a mean dependence on $M_\mathrm{h}$ (\citealt{dutton_2014} but see Eqn~2 from \citealt{posti_2018a}) and with a scatter of 0.11 dex. 
\item Compute $j_\mathrm{h}$ using Eqn~\ref{eqn:jh}.
\item Draw $j_*$ from a normal distribution in $\log_{10}j_*$ with a mean given by the \jm\ relation shown in panel E of Figure~\ref{fig:j-M-mueff} for the \mueff-selected subset of $N=694$ galaxies with $\alpha=0.57$ and $\beta=-2.55$.   A normally-distributed random scatter of standard deviation 0.083~dex (i.e., the intrinsic scatter of the \jm\ relation being used) is added. 
\item Calculate $f_j=f_*/f_\mathrm{h}$.  

The results from this process are shown in Figure~\ref{fig:fj_Vs_Mstar}.  Clearly, the mean SHSAMR (thick grey curve) is made up of two exponential components separated at $\log_{10}(M_*/M_{\odot})\sim~10.4$\footnote{For comparison, the best-fit characteristic mass used in Eqn~\ref{moster} was $\log_{10}(M_1/M_{\odot})=11.59$}.  The thin grey curves above and below the mean relation represent the standard deviations of Gaussians fit to the vertical $\log_{10}f_j$ slices.  The typical Gaussian scatter about the mean relation has $\sigma\sim 0.4$~dex.  Given that the uncertainties of the various quantities used to generate $f_j$ are certainly not all unrelated to one another, the mean error of $\sim 0.4$~dex on $f_j$ can safely be regarded as a conservative upper limit.  The low intrinsic scatter ($\sigma_\mathrm{int}=0.083$) of the \jm\ relation used to generate the SHSMAR noticeably lowers its scatter.  \jm\ relations from the literature oftentimes have associated scatters a factor 2 to 3 higher, which increase the scatter in the SHMAR presented here by $\sim 25\%$.  

\begin{figure}
	\includegraphics[width=\columnwidth]{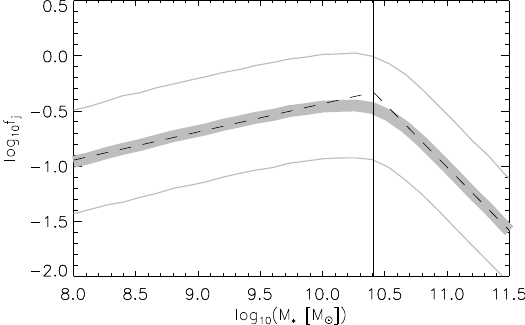}
    \caption{Fraction of stellar-to-halo specific angular momentum as a function of stellar mass.  The solid grey curve represents the mean relation, while the thin grey curves represent the $1\sigma$ spread in $\log_{10}f_j$ at a given $\log_{10}M_*$.  The relationship between $f_j$ and $M_*$ is well-modelled by two different power-laws (represented by the black-dashed lines).  For $\log_{10}(M_*/M_{\odot}) \lesssim 10.4$, $f_j\propto M_*^{0.25}$, while $f_j\propto M_*^{-1.15}$ above this mass threshold.  Tight \jm\ relations such as those presented in this study can be used to better constrain the uncertainty in the relation between $f_j$ and $M_*$. }
    \label{fig:fj_Vs_Mstar}
\end{figure}


The main aim of the experiment presented in this section is to very simply demonstrate just one way in which very tight empirical \jm\ relations (such as those presented in this work) can be used to effectively constrain the links between the measurable properties of galaxies and the properties of their dark matter haloes.

\end{enumerate}
\section{Summary}\label{sec:summary}
This study presents measurements of the empirical relation between stellar specific angular momentum ($j_*$) and stellar mass for large samples of ALFALFA galaxies.  Three different catalogues are used to gather the required  observed and derived galaxy properties.  A focus is placed on the disc components of galaxies (as modelled by an exponential light profile) that have small relative errors on their derived stellar masses and various other properties required to estimate $j_*$.  In total, $N=3~607$ galaxies are used in this study. The sample spans $\gtrsim 3$ decades in stellar mass and is highly diverse in terms of cold gas fraction of the interstellar medium.

The \jm\ relation based on all $N=3~607$ galaxies has $j_*$ well-related to stellar mass via a power-law index $\alpha=0.404\pm 0.003$ with an intrinsic scatter of 0.168 dex.  When considering the cold-gas fraction of interstellar medium, $f_\mathrm{gas}$, as a second parameter that controls a galaxy's $j_*$ content, it is shown that \fgas-selected sub-samples follow \jm\ relations that have higher power-law indices and which compare well to literature results based on high-quality rotation curves and spatially resolved imaging. However, the amount of intrinsic scatter in the \jm\ relations for the \fgas-selected sub-samples is not significantly lower than it is for the full galaxy sample, suggesting \fgas\ to be of limited importance in the context of a galaxy's $j_*$ content.  

Given the established importance of a galaxy's bulge-to-total mass fraction in determining its $j_*$ content, this study investigates the role played by a system's mean $I$-band surface brightness within its effective radius, \mueff.  A very tight plane can be fit to the 3D distribution of galaxies in $\log_{10}j_*$~--~$\log_{10}M_*$~--~\mueff\ space.  When splitting the ALFALFA galaxies into \mueff-selected sub-samples, extremely tight \jm\ relations with intrinsic scatters as low as 0.083~dex are generated for intermediate-mass galaxies with \mueff\ in the range 24~--~20~mag~arcsec$^{-2}$.  These are some of the very tightest \jm\ relations produced to date.  Furthermore, most of them are based on galaxy samples that are several factors larger than any other samples presented in the literature. The dependence of a galaxy's deviation from the mean \jm\ relation on its \mueff\ is explicitly demonstrated.  At a particular $M_*$, a galaxy with high/low stellar flux per unit area within its effective radius will have less/more than the average amount of stellar specific angular momentum of galaxies at that mass.  

To demonstrate the utility of a tightly constrained \jm\ relation, a method of constraining the stellar-to-halo specific angular momentum relation is presented.  Using random samplings of various dark matter halo parameters as well as random $j_*$  samplings based on the tightest \jm\ relation presented in this study, the $M_*$ dependence of the ratio of stellar-to-halo specific angular momentum is presented, and is shown to vary significantly with stellar mass.  While the form of this relation is largely determined by the choice of stellar-to-halo mass relation, it is shown that its scatter can be noticeably reduced by using tight \jm\ relations such as those presented in this work, thereby demonstrating a manner in which the empirical \jm\ relation can be used to link the observable properties of galaxies to their dark matter halo properties. 

The \jm\ relations presented in this work are among the tightest, most accurate ever measured, and are based on samples that are significantly larger than any others considered before.
\section{Acknowledgements}
Sincere thanks are extended to the anonymous referee for offering truly insightful and constructive comments that ultimately served to boost the quality of this work.

%


\section*{Data Availability}
All catalogues used in this study are publicly available.  Upon reasonable request, the author is willing to make available the derived specific angular momentum quantities.



\bibliographystyle{mnras}
\bibliography{bibliography} %


%


\bsp	
\label{lastpage}
\end{document}